\documentclass[12pt]{article}
\usepackage{authblk}
\usepackage{amsmath}
\usepackage{float}

\usepackage{graphicx}
\usepackage{dcolumn}
\usepackage{bm}

\usepackage[natbib=true,style=numeric,sorting=none]{biblatex}
\usepackage[right=2.5cm, bottom=2.5cm, left=2.5cm, top=2cm]{geometry}

\usepackage{xcolor}

\usepackage{lineno}

\begin{document}

\title{Total Ionizing Dose Measurements in Small Satellites in LEO using LabOSat-01}

\author[1]{\small Lucas Finazzi\footnote{Lead author: lfinazzi@unsam.edu.ar}}
\author[1,2]{Mariano Barella}
\author[3]{Fernando Gomez Marlasca}
\author[4]{Lucas Sambuco Salomone}
\author[4]{Sebastián Carbonetto}
\author[4]{María Victoria Cassani}
\author[4]{Eduardo Redín}
\author[4]{Mariano García-Inza}
\author[1]{Gabriel Sanca\footnote{Corresponding author: gsanca@unsam.edu.ar}}
\author[1]{Federico Golmar}

\affil[1]{Instituto de Ciencias Físicas, Universidad de San Martín-CONICET, Buenos Aires, Argentina}
\affil[2]{Department of Physics, University of Fribourg, Fribourg, Switzerland}
\affil[3]{Comisión Nacional de Energía Atómica, Buenos Aires, Argentina}
\affil[4]{Instituto de Tecnologías y Ciencias de la Ingeniería, Universidad de Buenos Aires-CONICET, Buenos Aires, Argentina}

\date{\today}

\maketitle 

\begin{abstract}
    LabOSat-01 is a payload designed to perform characterization experiments on electronic devices in hostile environments. Both Commercial-Off-The-Shelf components and custom nano and micro devices were studied in the last decade with this platform. The Total Ionizing Dose (TID) received by small satellites in Low Earth Orbit was measured with LabOSat-01 using p-MOSFET COTS and the results are presented in this work. Measurements were performed on two satellite missions, each totalling 1100 days, launched three years apart. Dosimeters were integrated in different satellite positions, with different shielding, and the measured TID after 1100 days ranged from 5~Gy to 19~Gy.
\end{abstract}

\section{Introduction}

Electronics inside small satellites in Low Earth Orbit (LEO) are constantly subjected to radiation from various sources. The two most important contributions in this orbit to the received dose come from trapped electrons and protons in the Van Allen Belts~\cite{tid_leo}. Monitoring the Total Ionizing Dose (TID) on the electronics helps estimate the lifespan of certain critical components and provides information on future mission development. 

Metal Oxide Semiconductor (MOS) dosimeters are transistors that allow measurement of the TID through the shift of its threshold voltage ($V_T$) caused by buildup of positive charge in the oxide and the generation of interface traps~\cite{traps_1, traps_2}. MOS dosimeters have already been used in space applications~\cite{space_app_1, space_app_2, space_app_3} and medical applications~\cite{medical_app_1, medical_app_2}. They are usually manufactured with gate oxides of hundreds of nanometers or even few micrometers, which makes them thicker than in regular MOS transistors~\cite{space_app_2, medical_app_1, oxide_thickness_1}. The increased oxide thickness for these types of devices stems from the fact that their sensitivity is approximately proportional to the square of their oxide thickness $t_{ox}$~\cite{oxide_thickness_2, oxide_thickness_3}. It is a well-known fact that these dosimeters can exhibit temperature dependence, can have high sensitivity dispersion (up to 30~\%) if manufactured in different Silicon wafers~\cite{FN_injection, si_wafer_1} and lose accumulated charge with time by a phenomenon called fading~\cite{fading_1}. This phenomenon is associated with the neutralization of oxide trapped charge via tunneling or thermal excitation of electrons~\cite{traps_1, traps_2}. All these limitations can be surmounted with a complete individual characterization before use.

A payload known as LabOSat-01 (\mbox{LS-01}) was designed to characterize and measure component behaviour in LEO~\cite{sanca2023}. \mbox{LS-01} underwent space qualification through exposure to thermal cycling, thermal shocks in a vacuum environment, mechanical shocks on a shaker, irradiation with high-energy protons, and thermal neutrons~\cite{tesisMariano}. \mbox{\mbox{LS-01}} was specifically designed to characterize two and three-terminal electronic devices under test (DUTs) in DC mode, functioning as a programmable Source-Measure Unit (SMU) capable of conducting I-V (current-voltage) curve measurements~\cite{barella2016}. For each unique DUT, the \mbox{LS-01} firmware is customized to execute a dedicated set of measurements and these DUTs are typically incorporated into \mbox{LS-01} using SOIC-16 packages. Additionally, \mbox{LS-01} includes an expansion connector that provides access to the SMU for external daughter boards.

Since 2014, we have successfully integrated nine payloads in eight different small satellites~\cite{9_missions_decade} operated by Satellogic~\cite{satellogicweb}. Each board remained operational throughout the life of its respective satellite mission. Currently, four of these systems are actively conducting experiments on DUTs operating in orbit, with regular data downloads for analysis. Some of the previously studied DUTs are RRAM devices~\cite{barella2019}, thin-film field-effect transistors~\cite{sanca2017} and Silicon Photomultipliers~\cite{barella2020}, among others. 

In this work, we present the results from two separate missions in which the \mbox{LS-01} platform was used to measure TID inside small satellites using MOS dosimeters. The first mission launched onboard ÑuSat-3 satellite (COSPAR-ID 2017-034C) on June 2017 and contained only dosimeters mounted (or integrated) directly into the \mbox{LS-01} platform. This mission lasted 1100~days. In addition, the second mission launched onboard the ÑuSat-7 (COSPAR-ID 2020-003B) on January 2020 and contained remote dosimeters placed along strategic points of the host satellite, which were connected to the \mbox{LS-01} platform via wires. This mission lasted 1100~days. In Section~\ref{sec:mat_and_meth}, we report on the MOS dosimeter characterization on Earth, the \mbox{\mbox{LS-01}} subsystem used to measure these dosimeters in LEO and their subsequent positions on the host satellite. In Section~\ref{sec:results}, we present the results obtained from both missions. In Section~\ref{sec:conclusions}, we present the conclusions of the work and discuss possible future improvements.

\section{Materials and Methods} \label{sec:mat_and_meth}

In this Section, the dosimeter characterization procedures and the results obtained from these are presented. The electronics used for LEO measurements are also detailed. 

\subsection{Dosimeter Characterization}

The dosimeters used in this work are COTS (Commercial off-the-shelf) p-channel MOS enhancement transistors with a Gate oxide of $t_{ox} \sim 250$~nm. The MOS devices are packaged in a TO-72 metal case. A representative subset of dosimeters from a larger batch was selected for individual characterization. All the selected devices had similar drain current to Gate voltage (i.e., similar $V_T$ and transconductance values).

During the characterization, MOSFET dosimeters were connected in diode configuration, where the Gate and Drain terminals and the Bulk and Source terminals are in short-circuit, respectively. The Source-Bulk pair is fixed at 0~V and a Gate-Drain voltage sweep was performed while the Drain current was measured. This was done to replicate the operating conditions in LEO (in the satellite, the dosimeters are irradiated continuously even when the electronics are off and these are only biased for readout). The Gate-Source voltage $V_{GS}$ shift is used to estimate the $V_T$ shift after irradiation. The following characteristics were obtained for the dosimeters under study:

\begin{itemize}
    \item Zero Temperature Coefficient (ZTC)
    \item Radiation Sensitivity
    \item Fading coefficient
\end{itemize}

\begin{figure}[!ht]
\centerline{\includegraphics[width=0.7\linewidth]{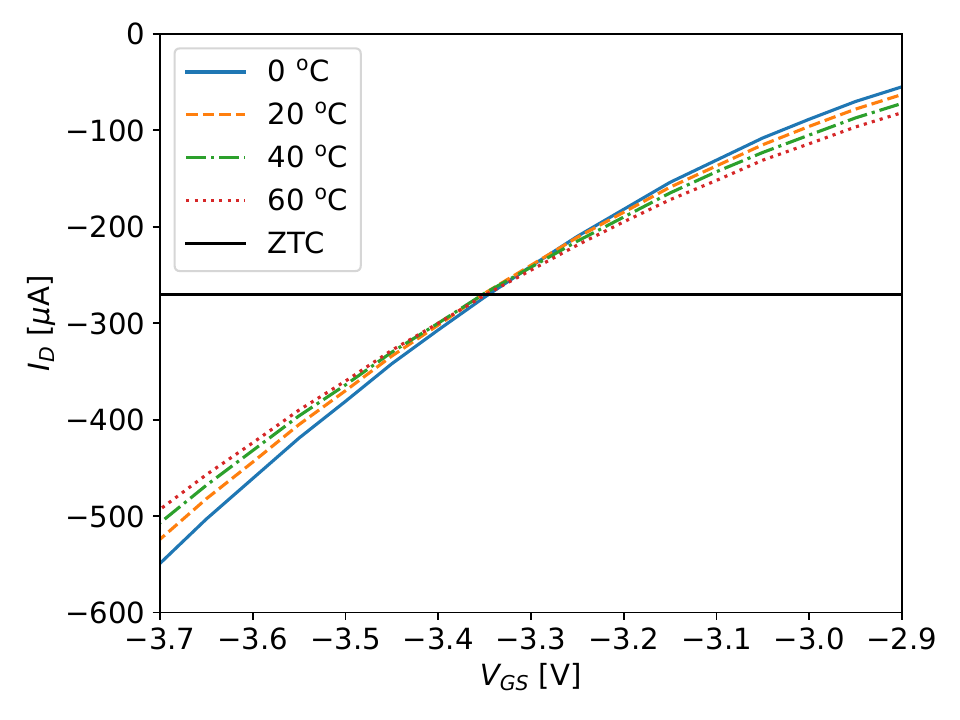}}
   \caption{I-V curves for a MOS dosimeter for 4 different temperatures along the calculated ZTC current of $(270 \pm 2)~\mu$A. The voltage value for this current is $(-3.35 \pm 0.01)$~V.}
   \label{fig:ZTC_point}
\end{figure}

Temperature variations affect the I-V characteristics of MOSFET devices. An increase in temperature leads to a decrease of both the absolute value of $V_T$ and the mobility of carriers in the conduction channel. Both effects tend to compensate for a given drain current $I_D$ value known as the Zero Temperature Coefficient (ZTC), which is the point in which all I-V curves intersect. Thus, to minimize temperature effects on dose measurements, the ZTC current $I_{ZTC}$ point is typically used to measure the $V_T$ shift in MOS dosimeters. Figure~\ref{fig:ZTC_point} shows I-V curves measured at different temperatures for a representative dosimeter. ZTC current (dark horizontal line) was found to be 270~$\mu$A. All dosimeters presented similar $I_{ZTC}$ values. 

To obtain the sensitivity, a $^{60}$Co source was used to irradiate each dosimeter with gamma photons. The I-V curve for each device was measured before and after irradiation. As previously mentioned, the Gate bias was biased at 0~V. Again, during the irradiation, the Gate was biased at 0~V to replicate operation conditions in LEO. The dosimeter batch for NuSat-3 mission was calibrated with a total dose of 11 Gy. The batch for NuSat-7 mission was calibrated with 1.9 Gy. The first batch was irradiated with a higher dose because these dosimeters had a higher $V_T$ than the other batch. It was believed a priori that the higher dose irradiation in these dosimeters would reduce the uncertainty in the sensitivity calculation. In the end, the uncertainties of all dosimeters are practically the same for different batches (2~mV/Gy).

\begin{figure}[!h]
\centerline{\includegraphics[width=0.7\linewidth]{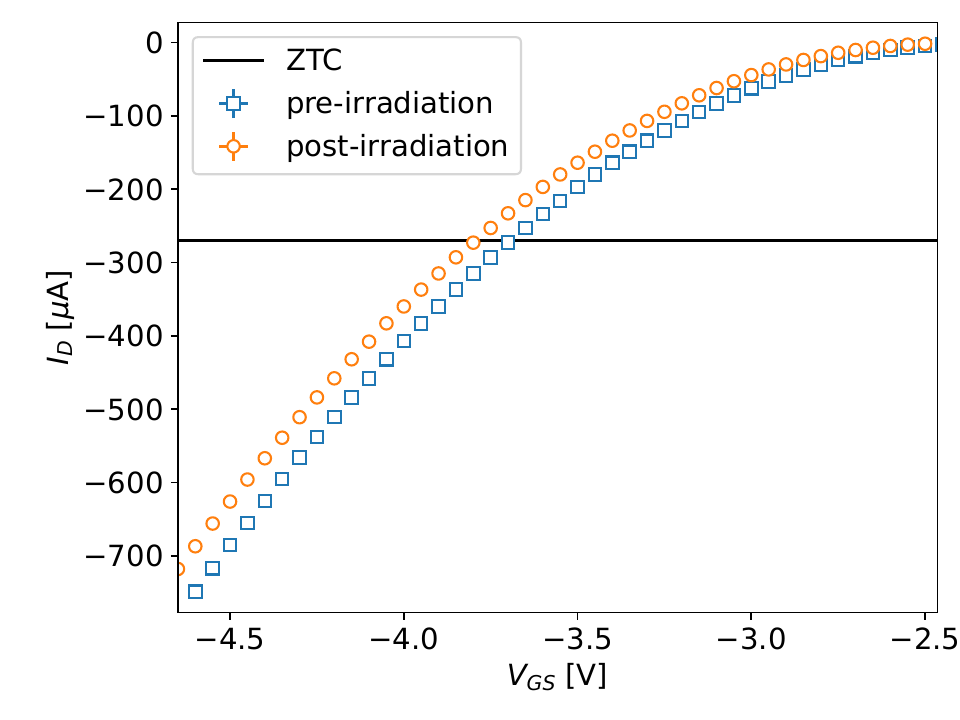}}
   \caption{I-V curve for a specific MOS dosimeter pre and post-irradiation. These curves are a way to measure the shift of $V_T$ for any operating drain current $I_D$. For this dosimeter, the sensitivity is $(10 \pm 2)$~mV/Gy. This dosimeter corresponds to the batch selected for the second mission. Error bars are small enough that they are not appreciated in this plot.}
   \label{fig:prepost_irradiation}
\end{figure}

An example of an I-V curve difference pre-irradiation and post-irradiation is shown in Figure~\ref{fig:prepost_irradiation}. This curve was measured with a Keithley~617. The systematic error for this device is 0.15\% of reading between 2~$\mu$A and 20~mA (current measurement) and 0.05\% of reading between 2~V and 20~V (voltage source).

The operating current was set to $I_{ZTC}$ and the sensitivity is calculated as the $V_{GS}$ shift for a given dose. This calibration is done this way because the $V_{GS}$ shift is expected to be linear for moderate doses as the ones expected in these Low Earth orbits.

Finally, the fading coefficients for the dosimeters were measured at room temperature. The fading coefficient $F$ can be defined after a time $t$ as 

\begin{equation}
    F(t) = \frac{V_T(\mathrm{ post-irradiation }) - V_T(\mathrm{ t })}{V_T(\mathrm{ post-irradiation }) - V_T(\mathrm{ pre-irradiation })} \ .
    \label{eq:fading}
\end{equation}

This coefficient is significant because the fading effect induces a drift in the dosimeter’s signal, attributed to a depletion of accumulated charge within the oxide layer. 

The fading coefficient for dosimeters from the first mission was estimated using 12 other dosimeters from the same batch. Due to the large fluctuation between different dosimeter fading coefficients, the uncertainty for this estimated value is large. For the second mission, the $V_T$ shift was measured 660~days after the initial irradiation and the fading coefficient was calculated for each dosimeter using Equation~(\ref{eq:fading}).

Table~\ref{tab:all} summarizes the characterized parameters of the dosimeters used in both missions. Mounted dosimeters from the first mission are labeled 1-3 and Remote dosimeters from the second mission are labeled 4-6 throughout the work.

\begin{table}[!h]
    \centering
    \setlength{\tabcolsep}{3pt}
    \begin{tabular}{|c|c|c|}
    \hline
    \begin{tabular}{@{}c@{}}Dosimeter \\ number \end{tabular} & Sensitivity [mV/Gy] & \begin{tabular}{@{}c@{}}F @ 24~$^{\mathrm{o}}$C \\ (660 days) [\%]\end{tabular} \\
    \hline
    \multicolumn{3}{|c|}{First mission: Mounted dosimeters}\\
    \hline
        1 & $(10 \pm 2)$ & $(30 \pm 15)$ \\
        2 & $(10 \pm 2)$ & $(30 \pm 15)$ \\
        3 & $(10 \pm 2)$ & $(30 \pm 15)$ \\
    \hline
    \multicolumn{3}{|c|}{Second mission: Remote dosimeters}\\
    \hline
        4 & $(9 \pm 2)$ & $(25 \pm 4)$\\
        5 & $(9 \pm 2)$ & $(36 \pm 5)$\\
        6 & $(8 \pm 2)$ & $(22 \pm 4)$\\
    \hline
    \end{tabular}
    \caption{Characterization parameters of dosimeters used in both missions presented in this work}
    \label{tab:all}
\end{table}

The fading curve has a $log(t)$ dependency~\cite{fading_1}, which means that most of the effect takes place at the start. It was estimated that the fading at 660~days is already 85 to 90~\% of the fading value at 1100~days. Furthermore, the temperature for all dosimeters is at least 15~$^\mathrm{o}$C lower than ambient temperature when in orbit\cite{9_missions_decade, cae2023}, which means that the fading parameter is expected to be even lower than the reported values, giving even better information retention for dose measurements.

\subsection{\mbox{LS-01} Electronics}

\begin{figure}[!ht]
\centerline{\includegraphics[width=0.7\linewidth]{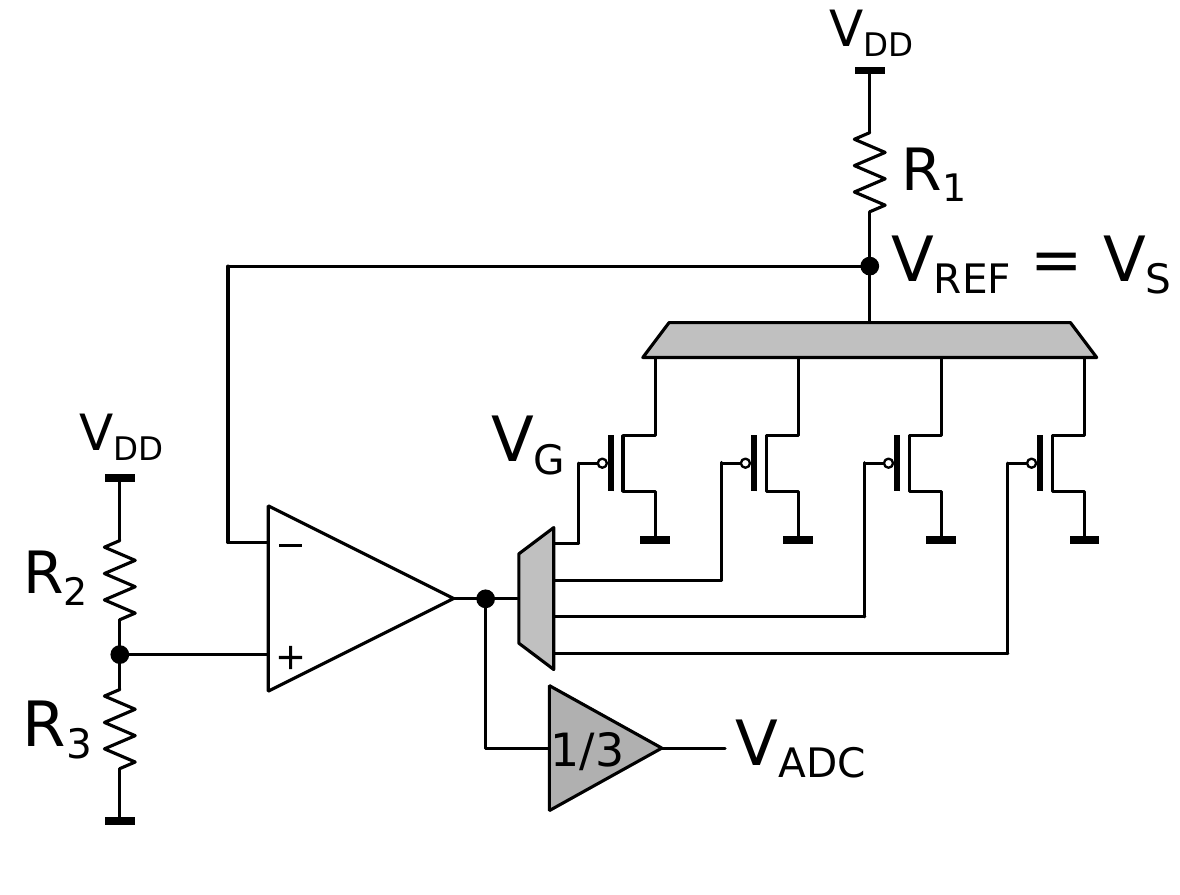}}
   \caption{Schematic circuit used to measure MOS dosimeters. Resistors are chosen specifically for each mission to obtain a drain current close to the ZTC current. Each dosimeter's Gate voltage $V_G$ is measured and $V_T$ shift was estimated indirectly as $V_{GS} = V_G - V_S$.}
   \label{fig:meas_circuit}
\end{figure}

To estimate the $V_T$ shift of MOS dosimeters with the \mbox{LS-01} board, a circuit that can guarantee a constant drain current is needed. In particular, to minimize temperature effects across all dosimeters, that current needs to be close to $I_{ZTC} = 270$~$\mu$A. In addition, the Gate voltage needs to be measured to estimate the $V_T$ shift of each dosimeter over time. This circuit is implemented in \mbox{LS-01} and a schematic of it is shown in Figure~\ref{fig:meas_circuit}. 

The schematic has the same topology for both missions but the value of $V_{DD}$ and the resistors (1~\% resistors were used) change depending on the mission. For the first mission, $V_{DD}$ is the satellite battery, which has a mean value of 12~V. Resistors were chosen so that $V_{REF} = (9.2 \pm 0.1)$~V. This means that the drain current over $R_1 = 10$~k$\Omega$ has a value of $I = (272 \pm 17)$~$\mu$A, which is sufficiently close to $I_{ZTC}$ to mitigate dosimeter temperature dependence.

Using the satellite battery voltage for $V_{DD}$ proved to be a problem during the first mission, as any fluctuation of this value instantly translated to bias current and gate voltage readout variations. To overcome this, the circuit was upgraded for the second mission by adding a voltage regulator that fixed $V_{DD} = 10$~V and adjusting the resistances to have $V_{REF} = (7.3 \pm 0.1)$~V and $I_D = (268 \pm 13)$~$\mu$A.

\mbox{LS-01} measures $V_{ADC}$ and $V_{GS}$ is estimated as follows:

\begin{equation}
    V_{GS} = V_G - V_S = 3 V_{ADC} - V_{REF} \ .
\end{equation}

Electronic switches are used so $V_G$ measurements are multiplexed and only one 12-bit Analog-to-Digital Converter (ADC) is needed for all integrated dosimeters. A 1/3 amplifier is used prior to the ADC due to the fact that the dosimeter Gate voltages are always higher than the maximum value of 2.5~V that the ADC can measure. The uncertainty in the value of $V_{GS}$ is mainly due to the contribution of the ADC quantization noise and from $V_{REF}$ fluctuations.

During both missions, the dosimeters are always off (0~V Gate bias and no Drain current) except when their Gate voltage is being read, which occurs at specific periods of time. To the effects of measuring TID, this ON time is negligible with respect to OFF time. Moreover, the circuit ensures that the dosimeter has zero Gate bias while OFF by placing a 1~M$\Omega$ pull-down resistor between the Gate and GND (not shown in Figure~\ref{fig:meas_circuit}). In addition to $V_{GS}$ measurements, temperature and battery voltage are measured during each experiment.

\begin{figure}[!h]
\centerline{\includegraphics[width=0.7\linewidth]{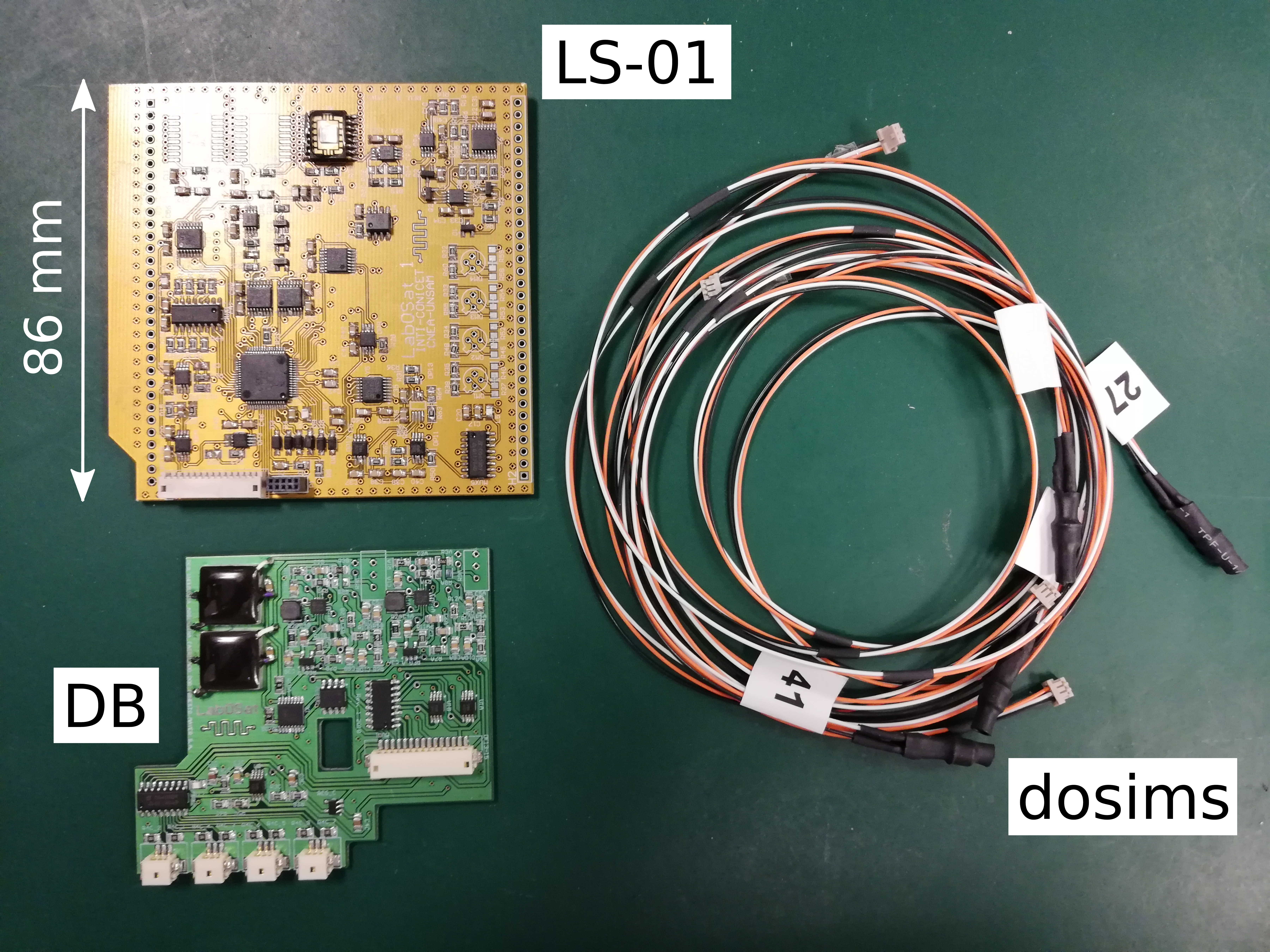}}
   \caption{Picture of \mbox{LS-01} board and Daughter Board (DB) used to connect the Remote dosimeters. This picture corresponds to the second mission outlined in this work.}
   \label{fig:electronics_pic}
\end{figure}

In Figure~\ref{fig:electronics_pic}, the electronics used for the second mission outlined in this work is shown. The Figure also shows the Remote dosimeters themselves ready to be integrated into the host satellite with long cable connections.

\subsection{Dosimeter Locations on Host Satellite}

\begin{figure}[!h]
\centerline{\includegraphics[width=0.7\linewidth]{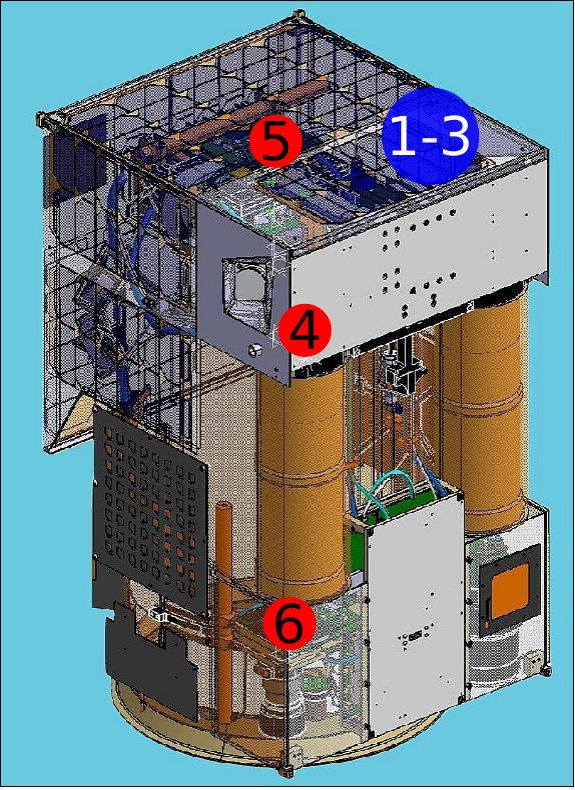}}
   \caption{Dosimeter positions inside host satellite. First mission Mounted dosimeters positions are shown in blue. Second mission Remote dosimeters positions are shown in red. Satellite's bottom points towards the Earth. Image adapted from~\cite{satellogic_satellite}.}
   \label{fig:dosim_positions}
\end{figure}

The type of satellite used to host the \mbox{LS-01} platform and the dosimeters is the same for both missions. The positions inside the satellite of the dosimeters considered in this work are shown in Figure~\ref{fig:dosim_positions}.

Mounted Dosimeters (1-3) and dosimeter 5 are placed on the top of the satellite and are more shielded against radiation than other parts because many satellite subsystems surround that region. Dosimeter 4 has less shielding than the previous dosimeters and dosimeter 6 is the most exposed one. The latter is located at the bottom of the satellite and it is next to two walls which have aluminum of $\sim$1.5~mm thickness. Besides, different dosimeters are subjected to different temperatures inside the satellite. In Table~\ref{tab:temps}, the temperature range of each dosimeter is shown. Mounted dosimeter temperature was measured with \mbox{LS-01} temperature sensor, while Remote dosimeter temperatures were obtained from satellite telemetry.

\begin{table}[!h]
    \centering
    \setlength{\tabcolsep}{3pt}
    \begin{tabular}{|c|c|}
    \hline
    Dosimeter number & Temperature range [$^{o}$C] \\
    \hline
    \multicolumn{2}{|c|}{First mission: Mounted dosimeters}\\
    \hline
        1 & $(0 \pm 5)$~$^{\mathrm{o}}$C \\
        2 & $(0 \pm 5)$~$^{\mathrm{o}}$C \\
        3 & $(0 \pm 5)$~$^{\mathrm{o}}$C \\
    \hline
    \multicolumn{2}{|c|}{Second mission: Remote dosimeters}\\
    \hline
        4 & $(-12 \pm 2)$~$^{\mathrm{o}}$C \\
        5 & $(5 \pm 10)$~$^{\mathrm{o}}$C \\
        6 & $(14 \pm 4)$~$^{\mathrm{o}}$C \\
    \hline
    \end{tabular}
    \caption{Temperature range of dosimeters for both missions during the whole mission duration.}
    \label{tab:temps}
\end{table}

\section{Results and Discussion} \label{sec:results}

\subsection{First mission: Mounted dosimeters}

\begin{figure}[!h]
\centerline{\includegraphics[width=0.7\linewidth]{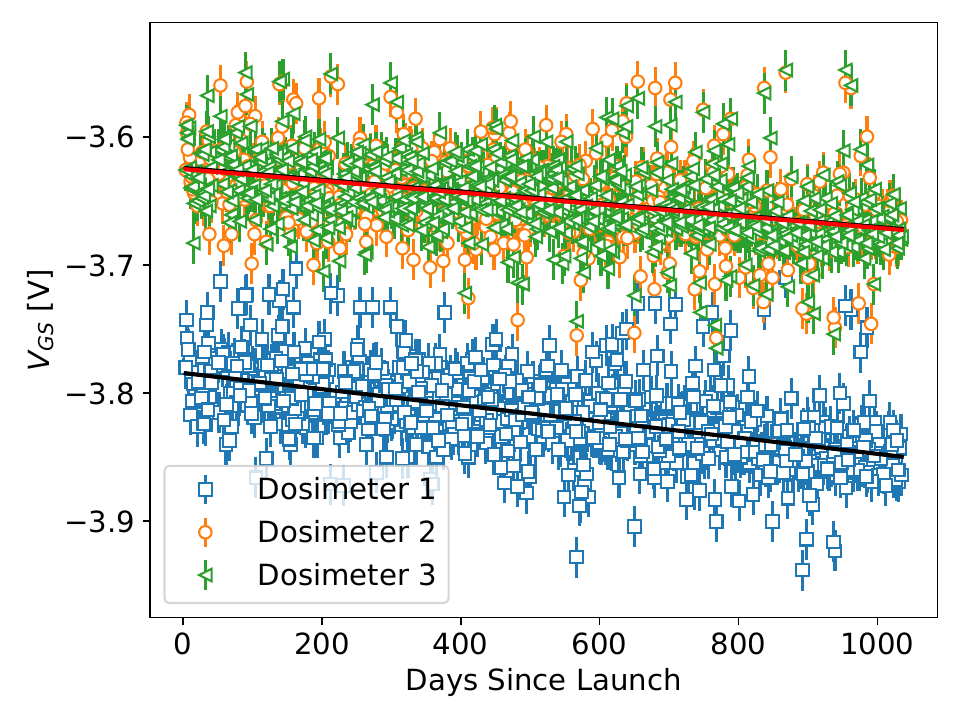}}
   \caption{$V_{GS}$ vs. days since launch for a total of 1100 days in orbit with corresponding linear fits used to estimate $V_{GS}$ shift. Reduced $\chi^2$ values for all fits were smaller than $10^{-3}$.}
   \label{fig:onboard_vgs_dsl}
\end{figure}

Figure~\ref{fig:onboard_vgs_dsl} shows $V_{GS}$ as a function of days since launch for the ÑuSat-3 mission (COSPAR ID 2017-034C), which lasted 1100 days, from June 15th 2017 to early 2020. As previously stated, in this mission the \mbox{LS-01} payload contained no voltage regulation for the battery voltage, so additional variation is expected in $V_{GS}$ measurements.

An overall mean $V_{GS}$ shift can be seen. Fast variations are observed due to a lack of supply voltage regulation. Dosimeter 1 had different initial $V_{GS}$ but shows a similar behaviour. 

\begin{figure}[!h]
\centerline{\includegraphics[width=0.7\linewidth]{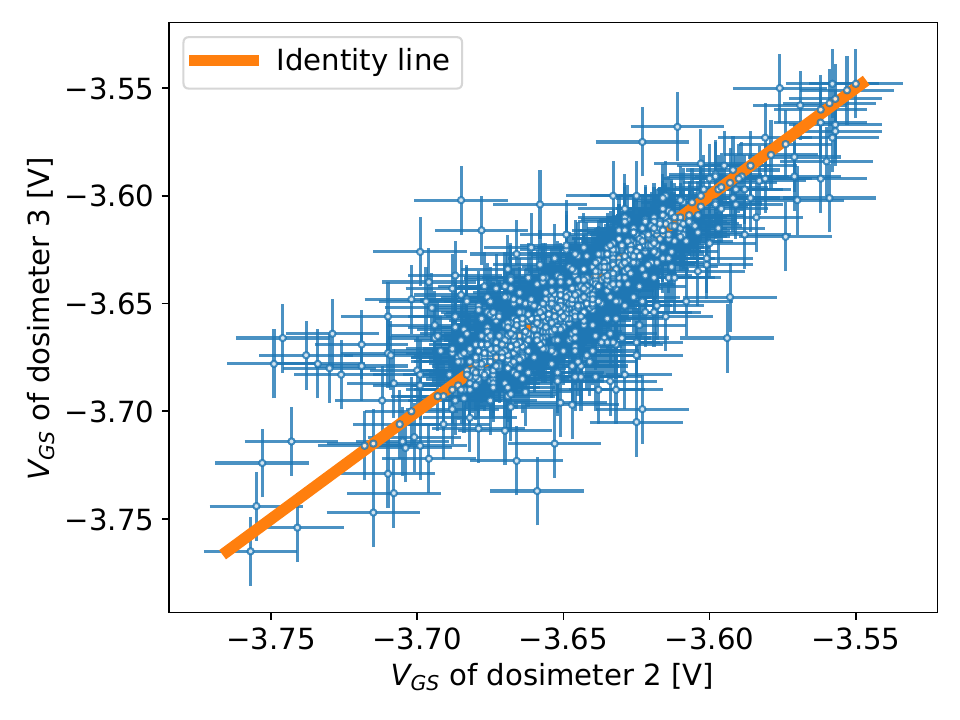}}
   \caption{Correlation plot between the $V_{GS}$ of dosimeter 2 and 3. An identity line is shown along with the $V_{GS}$ measurements. A strong correlation can be observed. The Pearson Correlation Coefficient for this dataset is 0.82. Such correlation is observed for all pairs of dosimeters but only one is shown for clarity.}
   \label{fig:correlation}
\end{figure}

The three dosimeters are expected to sense similar doses, as they are placed somewhat close together in the host satellite and have the same sensitivity. To validate this hypothesis, $V_{GS}$ vs. $V_{GS}$ curves for each pair of dosimeters were studied. Figure~\ref{fig:correlation} shows such plot for dosimeters 2 and 3. The distribution of points gathers around the identity line, suggesting a strong correlation.

To obtain the accumulated dose from each dosimeter, $V_{GS}$ measurements vs. time were divided by the appropriate sensitivity and a linear fit was performed to estimate the average dose per unit time. Table~\ref{tab:m1_doses} shows the obtained dose and their uncertainties for each Mounted dosimeter.

\begin{table}[!h]
    \centering
    \setlength{\tabcolsep}{3pt}
    \begin{tabular}{|c|c|}
    \hline
    Dosimeter number & Accumulated dose [Gy] \\
    \hline
        1 & $(6.2 \pm 0.4)$ \\
        2 & $(4.9 \pm 0.4)$ \\
        3 & $(4.5 \pm 0.4)$ \\
    \hline
    \end{tabular}
    \caption{Accumulated dose for each Mounted dosimeter after a linear fit. Dose uncertainties were obtained by error propagation of the obtained linear fit parameter uncertainties.}
    \label{tab:m1_doses}
\end{table}

It can be observed that the dose measured by dosimeter 1 is slightly larger than the measurement of the other two. This could be due to two different things: Firstly, each dosimeter might have different effective shielding. This is because they are in the payload bay (rather than next to satellite walls, like the Remote dosimeters) and additional surrounding electronics might provide extra shielding depending on dosimeter position. Secondly, fading might affect each dosimeter in a different way (fading for each individual dosimeter was not measured for this batch).

It is important to note that the dosimeter doses reported in Table~\ref{tab:m1_doses} represent the minimum received dose since fading effects were not taken into account in this case. Nevertheless, a maximum received dose value can be estimated in many cases adjusting by fading, as it is done for the second mission in the following Section.

\subsection{Second mission: Remote dosimeters}

The second dosimeter measurements presented in this work are from a \mbox{LS-01} payload onboard the ÑuSat-7 satellite (COSPAR ID 2020-003B). The satellite was launched on January 20th 2020 and data was received up to January 2024. The initial satellite orbit had a 97.34$^{\mathrm{o}}$ inclination and an apogee/perigee of 490/476~km, meaning that it has a polar orbit. Three Remote dosimeters were placed in critical positions inside the satellite as shown in Figure~\ref{fig:dosim_positions}. 

\begin{figure}[!ht]
\centerline{\includegraphics[width=0.7\linewidth]{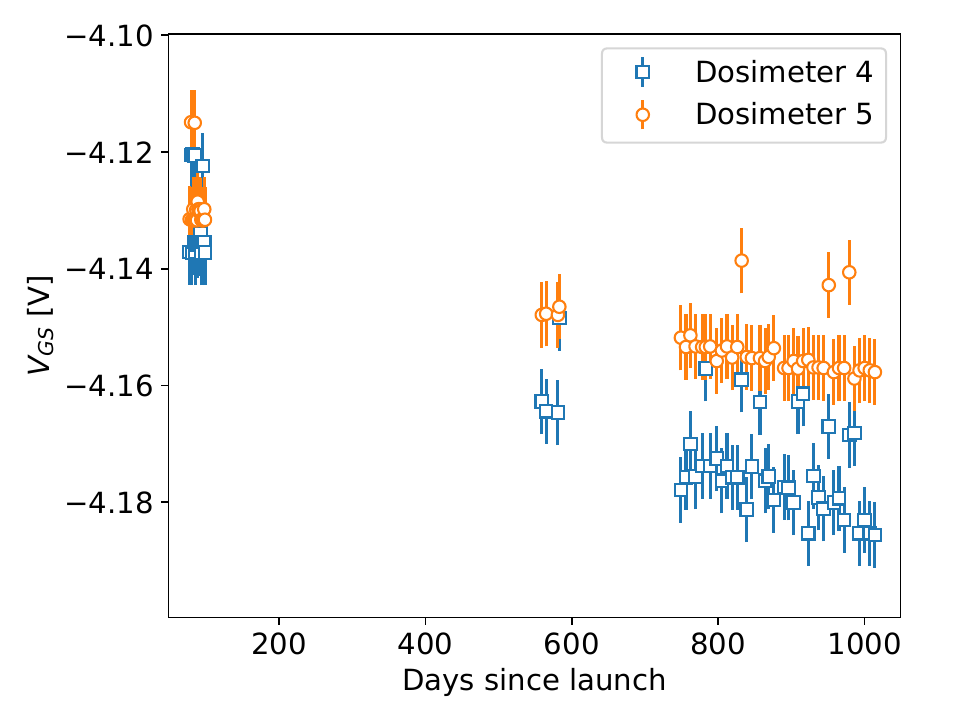}}
   \caption{$V_{GS}$ vs. days since launch for a total of 1100 days in orbit for two of the three Remote dosimeters. A similar shift in $V_{GS}$ can be seen when comparing the voltage shift from the previous mission. Dosimeter 6 was excluded from this plot due to clarity, as it had different initial $V_{GS}$.}
   \label{fig:m7_vgs_dsl}
\end{figure}

In Figure~\ref{fig:m7_vgs_dsl}, $V_{GS}$ as a function of days in orbit is shown.

After 1100 days, $\Delta V_{GS} \simeq -60$~mV, which is a similar value measured in the previous mission. As expected, $V_{GS}$ variations due to battery voltage were reduced, as now $V_{DD}$ is regulated.

\begin{figure}[!ht]
\centerline{\includegraphics[width=0.7\linewidth]{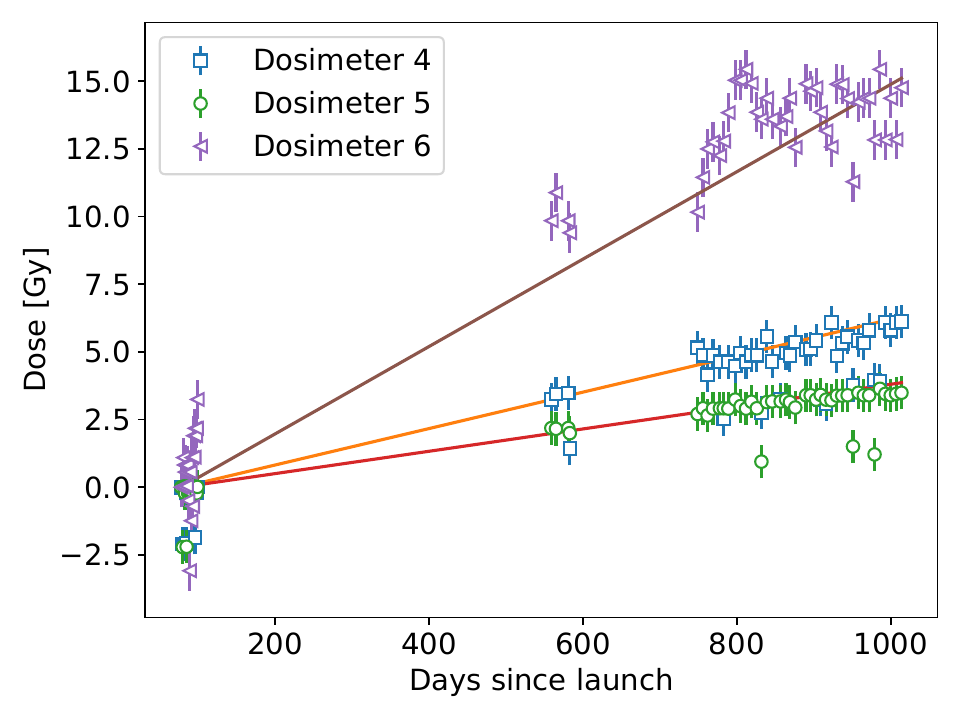}}
   \caption{Accumulated dose vs. days since launch for the 3 Remote dosimeters with their corresponding linear fits. Dose rates are different for each as they are placed in different places within the host satellite. }
   \label{fig:m7_dose_fits}
\end{figure}

The same procedure applied in the first mission is used to estimate the accumulated dose. Figure~\ref{fig:m7_dose_fits} shows the dose and their linear fits as a functions of time. 

Reduced $\chi^2$ for each fit performed are 2, 0.96, 3.9 for dosimeter 4, 5 and 6 respectively. The accumulated dose is measured from the first data point obtained, 70 days after mission launch. Notably, dosimeter 6 shows larger dose fluctuations than dosimeters 4 and 5, which in turn results in a large $\chi^2$ for the linear fit. This could be due to the fact that the ZTC point depends on the dosimeter's accumulated dose~\cite{ztc_variation}. More traps are created at the oxide interface with increasing dose and this effect is notorious for accumulated doses higher than $\sim$10~Gy. This effect results in a shift of the ZTC point and, in consequence, temperature dependant effects are no longer effectively compensated.

It is important to note that the presented doses in Figure~\ref{fig:m7_dose_fits} are the minimum received. In LEO, radiation is received in small doses over a prolonged time. This means that there are two competing phenomena in the dosimeters oxide at any given time: Radiation capture (or charge generation in the oxide) and neutralization (fading). Obtaining the actual dose received would require developing a fading model for the used dosimeters, but that is out of the scope of this work.

Nevertheless, a maximum bound for the dose received can be estimated if we take into account the good charge retention for these dosimeters. Fading values for these dosimeters in the host satellite are even lower than the values reported in Table~\ref{tab:all}, as they operate at lower temperatures. The benefit of having measured each fading individually is the much lower uncertainty when compared with the first mission dosimeters. This allows for a conservative estimate of a maximum bound on the received dose of each dosimeter separately, which is calculated as 

\begin{equation}
    \mathrm{Maximum \ Dose} = \frac{\mathrm{Minimum \ Dose}}{1 - F(\mathrm{660 \ days})} \ \ .
\end{equation}

The values obtained are presented in Table~\ref{tab:m7_doses}.

\begin{table}[!ht]
\setlength{\tabcolsep}{3pt}
    \centering
    \begin{tabular}{|c|c|c|}
         \hline
         Dosimeter number & Min. dose [Gy] & Est. max. dose [Gy] \\
         \hline
         4 & $(6.3 \pm 0.4)$ & $(8.4 \pm 0.7)$ \\
         \hline
         5 & $(3.9 \pm 0.3)$ & $(6.0 \pm 0.6)$ \\
         \hline
         6 & $(15.1 \pm 0.7)$ & $(19 \pm 1)$ \\
         \hline 
    \end{tabular}
    \caption{Dose measurements on the three Remote dosimeters and the estimated maximum dose received for each since data collection started.}
    \label{tab:m7_doses}
\end{table}

It can be seen that dose~5~$<$~dose~4~$<$~dose~6. This is expected, as dosimeter 6 is the one with less shielding, followed by dosimeter 4 and then by 5. To further validate these measurements, a \mbox{SPENVIS}~\cite{spenvis} simulation was carried out. Using the orbital parameters for this mission, a \mbox{SHIELDOSE-2}~\cite{shieldose} calculation was run to obtain the expected shielding thickness for each dosimeter for a given measured dose. The minimum, that is, the measured dose, was used for this. For the simulation, a silicon target was placed inside a spherical aluminum shell. The sphere-equivalent (or effective) shielding was determined to be 4.4~mm for dosimeter 4, 5.7~mm for dosimeter 5 and 2.9~mm for dosimeter 6. These values are comparable to those reported by the satellite manufacturer, which reported approximate equivalent shieldings of a couple of mm. Further details on the Remote dosimeter integration are protected by NDAs.

In addition, dose measurement values obtained are compatible with measurements reported in other works~\cite{results_comparison_1, results_comparison_2, results_comparison_3} for similar orbits and solar periods.

\section{Conclusions} \label{sec:conclusions}

The Total Ionizing Dose received by small satellites in Low Earth Orbit was measured with the LabOSat-01 payload using p-MOS COTS. These MOS dosimeters with $t_{ox} \sim 250$~nm were characterized and their ZTC current, sensitivity and fading was measured. It was observed that their information retention is very good even at 1100~days at near ambient Temperature. From these measurements, a value for the minimum received dose could be reported for both missions and a maximum dose bound could be estimated for the second mission. First mission dosimeter readings were validated observing correlations in $V_{GS}$ shifts for different dosimeters and second mission dosimeters were further validated by observing the effective shielding needed for the observed dose. These values are compatible with satellite specifications and with previous works that report measured dose inside satellites operating in LEO. Dose ranges are extremely useful for further mission design and component selection for future mission planning in similar LEO orbits.

In a future work, a detailed study on dosimeter fading will be performed. This will include a model based on physical principles, with the aim of obtaining an exact value of received dose for dosimeters~\cite{future_fading} irradiated with small doses over a prolonged time. In addition, a method will be developed to adjust the ZTC current as a function of the received dose to avoid unwanted measurement dispersion due to temperature effects.

\section*{Acknowledgements}

The authors would like to thank Satellogic for their help during the AIT, mission commissioning and operations. The authors acknowledge financial support from ANPCyT PICT 2017-0984 ``Componentes Electrónicos para Aplicaciones Satelitales (CEpAS)'', PICT-2019-2019-02993 ``LabOSat: desarrollo de un Instrumento detector de fotones individuales para aplicaciones espaciales'', UNSAM-ECyT FP-001, PICT 2020-1957, UBACYT: 20020220300077BA and 2012022041274BA.

\printbibliography

@Article{tid_leo,
    author = {Gutiérrez, O. and Prieto, M. and Perales-Eceiza, A. and Ravanbakhsh, A. and Basile, M. and Guzmán, D.},
    title = {{Toward the Use of Electronic Commercial Off-the-Shelf Devices in Space: Assessment of the True Radiation Environment in Low Earth Orbit (LEO)}},
    journal = {Electronics},
    volume = {12},
    year = {2023}
}

@ARTICLE{traps_1,
    author={Holmes-Siedle, A. G.},
    title={{The space-charge dosimeter: General principles of a new method of radiation detection},
    journal={Nuclear Instruments and Methods}},
    volume={121},
    year={1974},
    pages={169-179}
}

@ARTICLE{traps_2,
    author = {Holmes-Siedle, A. G. and Adams, L.},
    title = {{RADFET}: A review of the use of metal-oxide-silicon devices as integrating dosimeters},
    journal = {International Journal of Radiation Applications and Instrumentation. Part C. Radiation Physics and Chemistry},
    volume = {28},
    year = {1986},
    pages = {235-244}
}

@ARTICLE{space_app_1,
    author = {Adams, L. and Holmes-Siedle, A.},
    title = {{The Development of an MOS Dosimetry Unit for Use in Space}},
    journal = {IEEE Transactions on Nuclear Science},
    volume = {25},
    year = {1978},
    pages = {1607-1612}
}

@INPROCEEDINGS{space_app_2,
    author = {Holmes-Siedle, A. and Ravotti, F. and Glaser, M.},
    title = {{The Dosimetric Performance of RADFETs in Radiation Test Beams}},
    booktitle = {2007 IEEE Radiation Effects Data Workshop},
    year = {2007},
    pages = {42-57}
}

@ARTICLE{space_app_3,
    author = {MacKay, G.F. and Thomson, I. and Ng, A. and Sultan, N.},
    title = {{Applications of MOSFET dosimeters on MIR and BION satellites}},
    journal = {IEEE Transactions on Nuclear Science},
    volume = {44},
    year = {1997},
    pages = {2048-2051}
}

@ARTICLE{medical_app_1,
    author = {Rosenfeld, A. and Lerch, M. and Kron, T. and Brauer-Krisch, E. and Bravin, A. and Holmes-Siedle, A. and Allen, B.},
    title = {{Feasibility study of online high-spatial-resolution MOSFET dosimetry in static and pulsed X-ray radiation fields}},
    journal = {Faculty of Engineering - Papers},
    volume = {48},
    year = {2002},
    pages = {}
}

@ARTICLE{medical_app_2,
    author = {Rosenfeld, A.},
    title = {{Electronic dosimetry in radiation therapy}},
    journal = {Radiation Measurements},
    volume = {41},
    year = {2006},
    pages = {S134-S153}
}

@ARTICLE{oxide_thickness_1,
    author = {Lipovetzky, J. and {García Inza}, M. A. and Carbonetto, S. and Carra M. J. and Redin, E. and {Sambuco Salomone}, L. and Faigon A.},
    title = {{Field Oxide n-channel MOS Dosimeters Fabricated in CMOS Processes}},
    journal = {IEEE Transactions on Nuclear Science},
    volume = {60},
    year = {2013},
    pages = {4683-4691}
}

@ARTICLE{sanca2023,
  author={Sanca, G. A. and Barella, M. and Marlasca, F. Gomez and Alvarez, N. and Levy, P. and Golmar, F.},
  journal={IEEE Embedded Systems Letters}, 
  title={{LabOSat-01}: a payload for in-orbit device characterization}, 
  year={2023},
  volume={},
  number={},
  pages={1-1},
}

@BOOK{oxide_thickness_2,
    author = {Oldham, T. R.},
    title = {{Ionizing Radiation Effects in MOS Oxides}},
    publisher = {World Scientific Pub Co. Inc.},
    year = {1999},
}

@ARTICLE{oxide_thickness_3,
    author={Saks, N. S. and Ancona, M. G. and Modolo, J. A.},
    journal={IEEE Transactions on Nuclear Science}, 
    title={{Radiation Effects in {MOS} Capacitors with Very Thin Oxides at 80°K}}, 
    year={1984},
    volume={31},
    number={6},
    pages={1249-1255},
}

@ARTICLE{barella2020,
    author = {Barella, M. and Burroni, T. and Carsen, I and Far, M and {Ferreira Chase}, T. and Finazzi, L. and Golmar, F. and {Gomez Marlasca}, F. and Izraelevitch, F. and Levy, P. and Sanca, G.},
    title = {{Silicon photomultiplier characterization on board a satellite in Low Earth Orbit}},
    journal = {Nuclear Instruments and Methods in Physics Research Section A: Accelerators, Spectrometers, Detectors and Associated Equipment},
    volume = {979},
    year = {2020},
    pages = {164490}
}

@MASTERSTHESIS{tesisMariano,
    author = {Barella, M.},
    title = {{Dispositivos de memoria basados en {TiO2}: fabricación y caracterización en ambientes hostiles mediante un controlador dedicado}},
    school = {Universidad Nacional de General San Martín},
    year = {2018}
}

@INPROCEEDINGS{barella2016,
    author = {{Barella}, M. and {Sanca}, G and {Gomez Marlasca}, F. and {Rodrıguez}, G. and {Martelliti}, D. and {Abanto}, L. and {Levy}, P. and {Golmar}, F.},
    title = {{LabOSat: Low} cost measurement platform designed for hazardous environments},
    booktitle = {Proc. of 2016 Seventh Argentine Conference on Embedded Systems (CASE)},
    year  = {2016},
    pages = {1-6},
    address = {Buenos Aires, Argentina}
}

@INPROCEEDINGS{9_missions_decade,
    author = {Sanca, G. and Barella, M. and {Gomez Marlasca}, F. and Levy, P. and Golmar, F.},
    title = {{LabOSat: nine missions in a decade}},
    year = {2022},
    booktitle = {Proc. of Congreso Argentino de Sistemas Embebidos (CASE)},
    address = {La Plata, Argentina},
    pages = {}
}

@MISC{satellogicweb,
    author = {Satellogic},
    title = {{Creating a searchable Earth}},
    url = {http://www.satellogic.com/},
    urldate = {10/23/23},
    year = {2023}
}

@ARTICLE{barella2019,
    author = {Barella, M. and Sanca, G. and {Gomez Marlasca}, F. and {Román Acevedo}, W. and Rubi, D. and {García Inza}, M. A. and Levy, P. and Golmar, F.},
    title = {{Studying {ReRAM} devices at Low Earth Orbits using the LabOSat platform}},
    journal = {Radiation Physics and Chemistry},
    volume = {154},
    year = {2019},
    pages = {85-90},
}

@INPROCEEDINGS{sanca2017,
    author = {Sanca, G. A. and Barella, M. and Gomez Marlasca, F. and Rodr\'iguez,
    G. and Martelliti, D. and Patrone, L. and Levy, P. and Golmar, F.},
    title = {{LabOSat as a versatile payload for small satellites: first 100 days
    in LEO orbit}},
    booktitle = {{Proc. of the 1st Latin American Symposium of the International
    Academy of Astronautics on Small Satellites: Advanced Technologies and Segmented Architectures}},
    year = {2017},
    volume = {1},
    address = {Buenos Aires, Argentina},
    pages = {}
}

@ARTICLE{FN_injection,
  author = {Lipovetzky, J. and {Gabriel Redin}, E. and Faigon, A.},
  title = {Electrically Erasable Metal–Oxide Semiconductor Dosimeters}, 
  journal = {IEEE Transactions on Nuclear Science}, 
  year = {2007},
  volume = {54},
  pages = {1244-1250}
}

@ARTICLE{si_wafer_1,
  author={Faigon, Adrián and Lipovetzky, José and Redin, E. and Krusczenski, Gonzalo},
  journal={IEEE Transactions on Nuclear Science}, 
  title={Extension of the Measurement Range of {MOS} Dosimeters Using Radiation Induced Charge Neutralization}, 
  year={2008},
  volume={55},
  number={4},
  pages={2141-2147},
  doi={10.1109/TNS.2008.2000767}
}

@INPROCEEDINGS{fading_1,
    author = {Mousoulis, C and Elmiger, C. and Singhal, M. and Xuan, Y. and McNamee, T. and Thistlethwaite, J. and Walerow, P. A. and Salasky, M. and Scott, S. and Valentino, D. J. and Peroulis, D.},
    title = {{Characterization of fading of a MOS-based sensor for occupational radiation dosimetry}},
    booktitle = {2016 IEEE SENSORS},
    year = {2016},
    pages = {1-3}
}

@MISC{spenvis,
    url = {https://www.spenvis.oma.be/},
    title = {{ESA}, {SPace ENVironment Information System (SPENVIS)}},
    year = {2022},
    urldate = {02/01/24}
}

@MISC{shieldose,
    url= {https://nom.esa.int/models/sd2},
    title = {{SHIELDOSE-2}},
    urldate = {02/01/24},
    year = {2024}
}

@INPROCEEDINGS{results_comparison_1,
    author={Alvarez, Maite and Manzano, P. and Escribano, D. and Hernando, C. and Jimenez, J. J. and Sampedro, S. and Arruego, I.},
    booktitle={2016 IEEE Radiation Effects Data Workshop (REDW)}, 
    title={{On-Orbit measurements of {TID} and Dose Rate from two {RADFETs} on board {NANOSAT-1B} satellite}}, 
    year={2016},
    volume={},
    number={},
    pages={1-4},
}

@ARTICLE{results_comparison_2,
    author={Underwood, C.I.},
    journal={IEEE Transactions on Nuclear Science}, 
    title={{The single-event-effect behaviour of commercial-off-the-shelf memory devices-A decade in low-Earth orbit}}, 
    year={1998},
    volume={45},
    number={3},
    pages={1450-1457},
}

@inproceedings{results_comparison_3,
    title={{Shields-1 Dosimetry Measurements in Polar Low Earth Orbit}},
    author={Thomsen III, D Laurence},
    booktitle={NASA Small Spacecraft Systems Virtual Institute, Community of Practice Webinar Series},
    year={2022}
}

@misc{satellogic_satellite,
    url = {https://www.eoportal.org/satellite-missions/satellogic#spacecraft},
    title = {{{NewSat} (Aleph-1 Constellation)}},
    urldate = {2024/02/07}
}

@ARTICLE{ztc_variation,
  author={Carbonetto, Sebastián H. and Garcia Inza, Mariano A. and Lipovetzky, José and Redin, Eduardo G. and Sambuco Salomone, Lucas and Faigon, Adrián},
  journal={IEEE Transactions on Nuclear Science}, 
  title={{Zero Temperature Coefficient Bias in MOS Devices. Dependence on Interface Traps Density, Application to MOS Dosimetry}}, 
  year={2011},
  volume={58},
  number={6},
  pages={3348-3353},
}

@ARTICLE{future_fading,
  author={Sambuco Salomone, L. and Faigón, A. and Redin, E. G.},
  journal={IEEE Transactions on Nuclear Science}, 
  title={{Numerical Modeling of {MOS} Dosimeters Under Switched Bias Irradiations}}, 
  year={2015},
  volume={62},
  number={4},
  pages={1665-1673}
}

@INPROCEEDINGS{cae2023,
  author={Finazzi, L. and Sanca, G. A. and Marlasca, F. Gomez and Barella, M. and Izraelevitch, F. and Levy, P. and Golmar, F.},
  booktitle={2023 Argentine Conference on Electronics (CAE)}, 
  title={Study of Silicon Photomultipliers in Low Earth Orbit}, 
  year={2023},
  volume={},
  number={},
  pages={63-68},
  doi={10.1109/CAE56623.2023.10086977}
}


\end{document}